\documentclass[smallextended]{article} 
\usepackage{graphicx}
\usepackage{amsfonts} 
\usepackage[affil-it]{authblk}
\usepackage{lineno,hyperref}
\def\makeheadbox{\relax}

\begin{document}

\title{Lesion Conditional Image Generation for Improved Segmentation of Intracranial Hemorrhage  from CT Images \footnote{Accepted to SPIE Medical Imaging 2020.}}

\renewcommand\Affilfont{\fontsize{11}{10.8}\itshape}

\author{Manohar Karki}
\author{Junghwan Cho\footnote{Corresponding author. Email: jcho@caidesystems.com}
}
\author{Seokhwan Ko
}
\affil{CAIDE Systems Inc., Lowell, MA, USA}


%\titlerunning{Lesion Conditional Image Generation}
\date{\vspace{-5ex}}

\maketitle

\begin{abstract}

Data augmentation can effectively resolve a scarcity of images when training machine-learning algorithms. It can make them more robust to unseen images.  We present a lesion conditional Generative Adversarial Network (\textit{LcGAN}) to generate synthetic Computed Tomography (CT) images for data augmentation. A lesion conditional image (segmented mask) is an input to both the generator and the discriminator of the \textit{LcGAN} during training. The trained model generates contextual CT images based on input masks. We quantify the quality of the images by using a fully convolutional network (FCN) score and blurriness. We also train another classification network to select better synthetic images. These synthetic CT images are then augmented to our hemorrhagic lesion segmentation network. By applying this augmentation method on 2.5\%, 10\% and 25\% of original data, segmentation improved by 12.8\%, 6\% and 1.6\% respectively. 
\end{abstract}

%\begin{keyword}
%\texttt  {Data Augmentation \sep Conditional GAN \sep Lesion Segmentation \sep CT Image Generation}

%\end{keyword}

\section{Introduction}

Although deep learning architectures have solved challenging computer vision tasks in recent years \cite{gan}, \cite{krizhevsky2012imagenet}, \cite{szegedy2015going}, they require large amounts of data. In the medical field, collecting this vast amount of data is still quite challenging, and models tend to overfit if trained with limited data. As a solution to this problem, synthetic data is commonly added. Standard image transformation techniques like rotations, rescaling and contrast changes are some traditional methods of augmenting image datasets. These methods provide some variations in the dataset when there are a small number of samples. Nonetheless, these methods are still limited \cite{Hüttenrauch2016}, as each new synthetic image is a transformation of a single image.
\par
However, images that are synthesized using generative methods can extract information from the entire dataset. These methods learn the underlying distribution of the dataset \cite{gan}, \cite{kingma2013auto}. Each synthetic image is drawn from the distribution, rather than being transformations of a single image. The development of the Generative Adversarial Networks (GANs) has brought about practical usefulness of deep generative models  \cite{gan}. GANs have been used for style transfer  \cite{zhu2017unpaired}, texture synthesis \cite{dcgan}  and image-to-image transformations \cite{pix2pix}. GANs are comprised of two competing networks involved in a minimax game \cite{gan}. While the \textit{generator} produces data by estimating the probability distribution, the \textit{discriminator}  decides if the data is generated or original. Conditional generative adversarial networks are an extension of this approach  \cite{mirza2014conditional}. They attempt to learn multi-modal distributions with extra information, which is provided to both the generator and the discriminator. This is a flexible solution when data with specific attributes are desired.    

Our contributions in this paper are as follows:
\begin{itemize}
  \item We present a lesion-conditional GAN to synthesize Computed Tomography (CT) images. Hemorrhagic segmented lesion, which is an input condition to the network, and the generated CT images together form a CT image label mask pair.
\item  For the image generation model, we employ three criteria for evaluating image quality. These are: 1) how well these images can be segmented by a relevant segmentation model, 2) their visual quality, and 3) their perceived authenticity.  
  \item We examine the benefit of data augmentation with such synthetic images and segmentation label masks for a hemorrhagic segmentation network.
\end{itemize}
Deep learning algorithms for semantic segmentation also require ample amount of ground truth segmented labels (images). In this respect, our approach conveniently generates a tuple of CT images and ground truth masks from limited data.  Just like traditional augmentation, the CT image-label image pair can be used with the original data without further processing. In general, a random vector is used to generate data by a GAN. However, there is more control over the type of generated data when a segmented mask is used. The performance of artificially limited amounts is evaluated by mimicking the effectiveness of our method on datasets of variable sizes.

\section{Related Work}
 Algorithmic design choices while training models have been prevalent to regularize training. Some common methods include: in-network regularization, specialized layers for augmentation, appropriate penalty to the objective functions \cite{goodfellow2016deep}. Deep learning algorithms learn better fitting models when there are large amounts of diverse data\cite{cho2015much}. However, when data available is limited, data augmentation techniques can help improve performance. 
 \par
Augmentation has been widely used in machine learning algorithms especially for image-based datasets \cite{witten2016data}. While studying the effectiveness of data augmentation,  the authors of \cite{effective} analyze various data augmentation methods for image classification. They train an augmentation network using concatenated images of the same class. They also compare their approach with images generated using a GAN and traditional methods. For medical images, a generative solution \cite{lever} was proposed for the classification of liver lesion. The paper studied the usage of classical data augmentation followed by a GAN based approach to generate synthetic dataset. They also trained a separate GAN network for each class. 
\par 
U - Net \cite{ronneberger2015u}, which is a popular deep learning-based approach for semantic segmentation that is curated for biomedical image segmentation worked well primarily. The authors stress the need for data augmentation in  their approach. Since it is a supervised approach with an architecture built from encoder-decoder style predecessors, it still needs  relatively a large amount of labeled data to train.
There have been several other approaches for segmentation in medical images such as brain tumor segmentation \cite{havaei2017brain} and V-Net \cite{milletari2016v} for volumetric segmentation. These segmentation approaches still benefit from data augmentation.
Image style transfer \cite{gatys2016image} is a popular convolutional neural network approach that transfers known artistic style to a normal image. A solution with this method should be considered but it still requires images from \textit{normal} patients. Image pairs could be created by merging the segmentation mask (style) into such images, but some intervention may be needed for the image compositing.

\section{Method}
\label{sec:method}
Our proposed method employs generative adversarial networks (GANs) and takes masked lesion images as a condition to translate them into lesion corresponding CT images. Using the combination of generated images and some traditional transformations, a larger dataset is built. 
The segmentation architecture is a modified fully convolutional network (FCN-8s)  \cite{long2015fully}  suitable for this application.
\begin{figure}[]
\centering
\includegraphics[width=.9\textwidth]{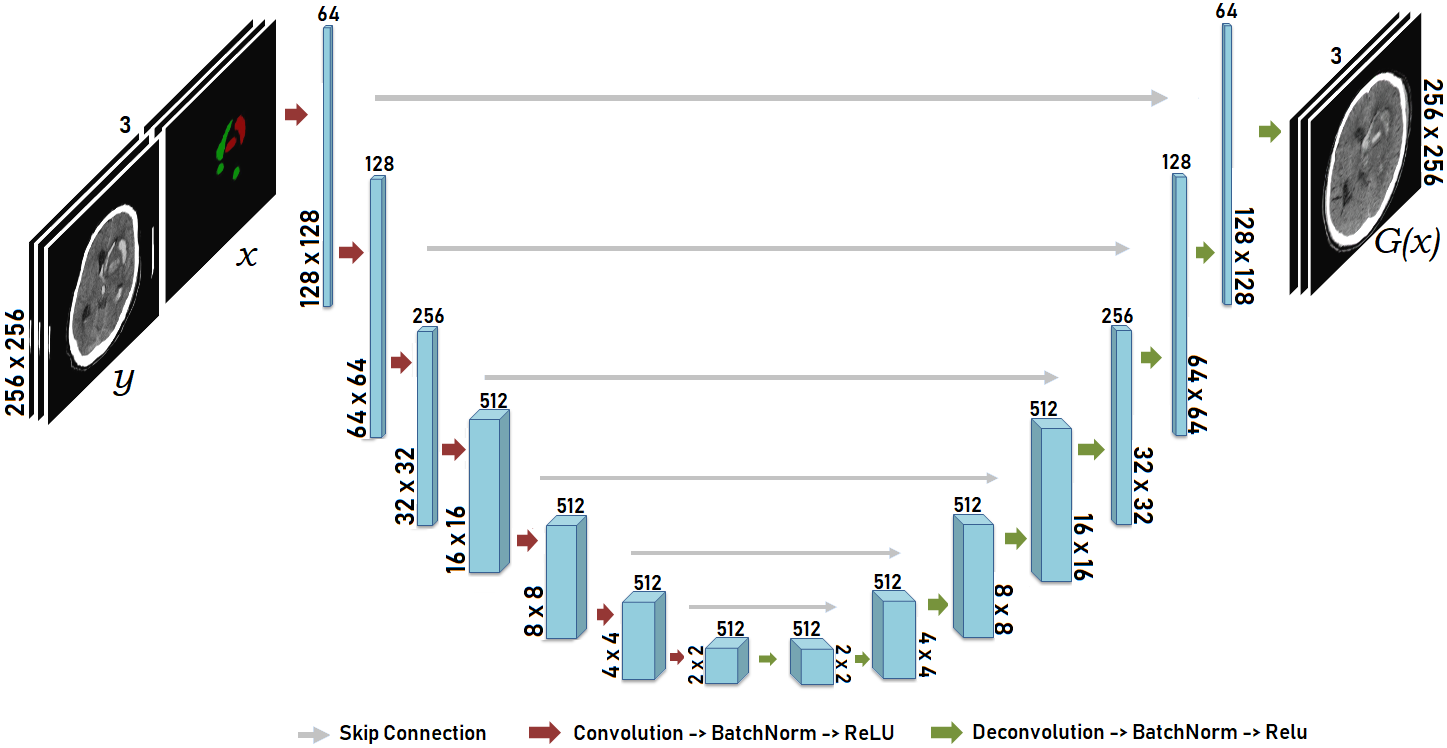}
\caption{The generator architecture is based on U-Net \cite{ronneberger2015u} which has skip connections between higher and lower levels. }
\label{gan_brain}
\end{figure}
\subsection{Conditional Generative Adversarial Networks}
\label{sec:ganarch}
 Conditional GANs (cGANs) learn the mapping of input $x$, along with a random vector $z$, to $y$. The generator learns to synthesize $G(x)$. The discriminator also \textit{sees} $x$ along with the generated data ($G(x)$) or the target data ($y$) and tries to discriminate the real data from the fake ones.  The objective loss function that a cGAN is trying to learn (from \cite{pix2pix}) is defined as follows :
 \begin {equation}
\mathcal{L}_{cGAN}(G,D) = \mathbb{E}_{x,y}[\log D(x,y)] +  \mathbb{E}_{x,z}[\log (1- D(x, G(x,z)))]
 \end {equation}

\begin{figure}[h]
\centering
\includegraphics[width=.65\textwidth]{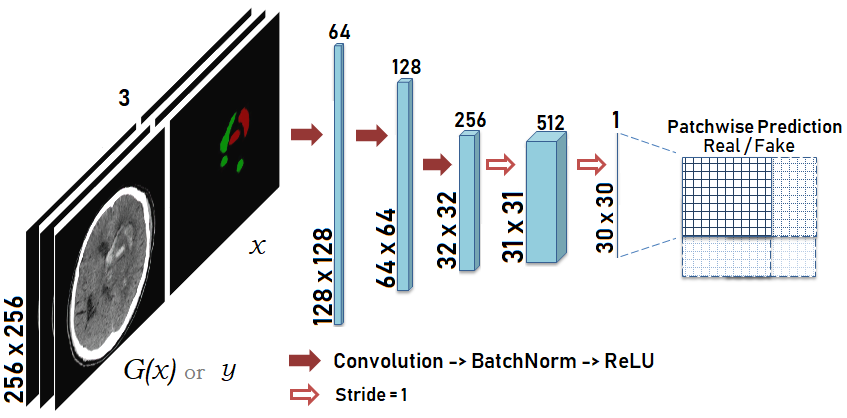}
\caption{Discriminator architecture is based on image to image translation \cite{pix2pix}. Final output image (30 x 30) corresponds to results from all patches (70 x 70) over the original image. }
\label{gan_brain2}
\end{figure}

When an image is the condition instead of a random vector $z$, the same input $x$ is shown to both the generator and the discriminator. As seen in Fig. \ref{gan_brain} and Fig. \ref{gan_brain2}, the same lesion segmented image is  an input to both the generator and discriminator.  The \textit{LcGAN} architecture follows the image to image translation approach where the discriminator makes decisions at the level of patches instead of pixels \cite{pix2pix}.  The generator has an architecture like U-Net with skip connections between convolution and deconvolution layers, and the discriminator is simply convolutional neural network (CNN) that classifies the patches as real or fake. The condition images are  segmented lesions, which are labeled for different types of hemorrhages. The segmented lesions help to steer the generation of CT images. The hemorrhage image ($y$) and its segmentation mask ($x$) are input to the generator during the training. The segmentation mask is also used as input to the discriminator along with the real CT image ($y$) or the generated CT image ($G(x)$) from the generator. The generator loss includes L1 loss since it has been known to yield less blurry images \cite{pix2pix}.  
\subsection{Image Quality Criteria}
The ultimate goal of generating synthetic images is to aid the training of the segmentation network. For this reason, \textit{LcGAN} models generating the best synthetic images must be chosen. GANs learn the loss function during the training which makes it impractical to determine the best fitting model. Different \textit{LcGAN} models with different hyperparameters yield with diverse qualities of images. The quality of images generated by such models can determine the best model.  Quality of generated images is difficult to quantify. Several methods have been proposed for this problem \cite{theis2015note}. However, since there is absence of universally accepted metric, we use the following metrics to measure the quality of images generated by \textit{LcGAN} models.
\label{sec:eval}

\subsubsection{FCN Score}
\begin{figure}[h]
\centering
\includegraphics[width=.8\textwidth]{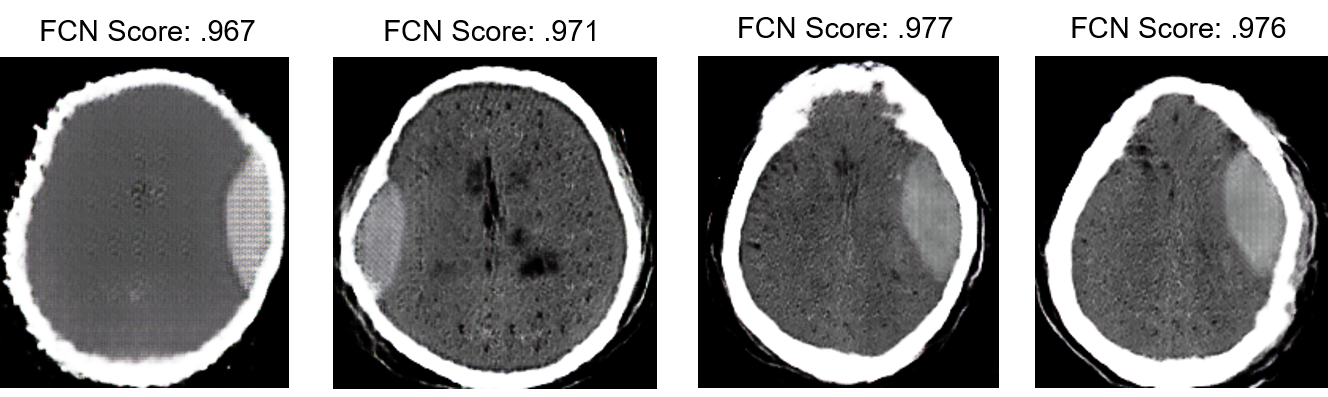}

\caption{Examples of generated images from different \textit{LcGAN} models using 10\% data with high FCN Scores all belonging to the same class (EDH). }
\label{fcn_scores}
\end{figure}

The most common way of evaluating images, which are generated using GANs, is to use another network trained on a similar dataset \cite{borji2018pros}. The classification performance of the generated images on the network is called \textit{Inception Score} and it represents the quality of the images. For segmentation tasks, it is more appropriate to use a segmentation performance value. When a fully convolutional network is used for the segmentation, the dice similarity coefficient can be used as metrics to  measure the quality of images.  However, upon inspection, images with comparable fully convolutional network (FCN) segmentation scores sometimes had very different quality. Fig. \ref{fcn_scores} shows examples of some images that have almost identical FCN scores, but the visual quality is quite different. While this is the most important metrics that eventually defines how our augmentation data improves our intended performance, we also present two other metrics to establish the quality of generated data.  

\subsubsection{Blurriness}
There are a variety of methods to determine how blurry an image is \cite{russ2016image}. Images can be converted to the frequency domain using fast Fourier transform to check the presence of high level features \cite{de2013image}. Blurry images will lack such features. The mean of all values  in the frequency domain is assigned as blurriness value for the image. Another option, we chose, is  the variance of Laplacian of each individual image \cite{pertuz2013analysis}. Laplacian is used for edge detection in images to determine vast intensity changes. Images that are not blurry will have sharp edges. For each of these methods, the mean blurriness of the entire set of images derived from a trained \textit{LcGAN} model can then be calculated. Both methods have drawbacks in determining blurriness in natural images. Natural images can be blurry locally in small regions and sharp in other areas. Brain CT images are more homogenous as they have salient pixels in the central region surrounded by background pixels (black) in the edges.  Hence, the comparison of blurriness calculated by these two approaches are meaningful.
	
\subsubsection{CNN based Classifier}
Some of the images, which had high FCN scores, were not very blurry but had unnatural patches. The  second image in Fig. \ref{fcn_scores} is an example of this.  Another convolutional neural network (CNN)  based classifier can act like a discriminator and choose realistic synthetic images. Using a combination of generated images with different parameters and the original images, a CNN  is trained to distinguish between the original and the generated images. While a probability of 50\% would be used for classification of two-class problem (real and fake), more fake images can be deemed acceptable by decreasing the threshold. Ideally, the best generative model should generate images that are harder to distinguish with this classifier. Models that can fool this classifier by generating the largest number of realistic images are chosen to generate the augmented data. 

%We calculate overall  recall, which is the average recall for the entire test set and  per class recall where we evaluate recall for each class over the entire test set. The following equation based on True Positive (TP), False Positive (FP) and False Negative (FN) is used to calculate recall:
% \begin {equation}
%Recall ={\frac {TP}{TP+FN}}
% \end {equation}
%Similar to per class recall, we calculate the per class DSC performance and is measured as follows: 
% \begin {equation}
%DSC_{class}={\frac {2TP}{2TP+FP+FN}}
% \end {equation}

\section{Experiment}
For our experiments, a NVIDIA Tesla V100 GPU  was used with TensorFlow and Caffe frameworks.  During these experiments, CT images consisting of at least one type of hemorrhage were included. There are 5 different types of intracranial hemorrhages that were segmented: 1) Intraparenchymal (IPH) 2) Intraventricular (IVH) 3) Subarachnoid (SAH) 4) Epidural (EDH) 5) Subdural (SDH). There was a total of 2117 patients' data in the training set and 660 patients' data in the testing set. Some patients had more images compared to others, and the data was selected based on patients. Each cross validation set still had similar number of image samples. Each set had 26k training and 6k validation hemorrhagic CT images. On the entire dataset, augmentation did not yield any improvements during the segmentation. We then limited the size of the training data from the original dataset and improved the performance of the limited dataset with augmentation.
 \par The synthetic image generation method using cGAN is developed from the TensorFlow adaptation \cite{affine} of the image to image translation implementation  \cite{pix2pix}. Only the synthetic images created from the constrained dataset were used to augment the constrained dataset. The CNN used for determining the quality of synthetic images was based on GoogleNet architecture \cite{szegedy2015going}. Original images and equal numbers of generated images from each of the trained models were used for training and validation of the network.
\subsection{Training the fully convolutional network}
\label{sec:fcnarch}
In segmenting the hemorrhagic lesions on the CT images, a fully convolutional network (FCN) was used. By replacing the fully connected layers of a conventional convolutional network, fully convolutional networks were designed, and they have been widely used to segment regions in images \cite{long2015fully}. The network consists of an encoder-decoder style architecture where segmentation mask is the final output. Our FCN uses the weights from a pretrained VGG-16 model \cite{simonyan2014very}. Augmented data  is fed into this network along with the original data in order to improve the segmentation.

\paragraph{Dice Similarity Coefficient:}
Segmentation by FCN was evaluated by using the Dice Similarity Coefficient (DSC). Using the original validation set,  each of the models trained with various percentages (2.5\%, 10\%, 25\%) of the total images was evaluated.  A large part of the images consists of background pixels. The inclusion of the background class during calculation is not very useful in measuring subtle improvements in performance. Hence, background pixels were ignored during the evaluation. 
	For the performance of segmentation, DSC metrics have been widely used, especially for medical images.  It is based on the ratio of overlapping area of the segmented output (S) and ground truth (G) to the total area of both. The DSC performance is calculated for each image and averaged over the entire evaluation set.
 \begin {equation}
DSC = \frac{2 (|S  \cap G| )}{ |S| +  |G| } 
 \end {equation}
\par The precision and recall of the entire test set are also evaluated by only accounting for each prediction that passes certain thresholds. These thresholds are based on the number of pixels in the image and DSC value. A false positive detection with less than 200 pixels area and a true positive detection where DSC score is less than .25 are not included. Instances where there were only a few pixels detected, were thus considered as  noise and thus, ignored.

\subsection{Training the lesion conditional GAN}
\label{sec:ganaug}
	The \textit{LcGAN} algorithm uses the segmentation masks of the original CT images to generate synthetic CT images. The size of the original dataset is increased to two-fold as every time each ground truth mask generates a new synthetic CT image based on it. During the \textit{LcGAN} training process, which follows the flow of image to image translation \cite{pix2pix}, hemorrhage segments were used for the inputs to both the generator for synthetic image generation and to the discriminator along with the real or fake image. After the training, the best models were chosen based on our criteria for image quality. With these models and label images from the training data, more data was generated.  They are then augmented to original data for training the segmentation algorithm. For training the \textit{LcGAN}, we used Adam solver and fixed our initial learning rate ($\alpha$) of .0002 and momentum ($\beta_1$) of .5 and trained for 10, 50, 100 and 200 epochs. The number of filters for each layer of the generator and discriminator is shown in Fig. \ref {gan_brain} and \ref{gan_brain2} respectively. These values are based on the image-image translation implementation in TensorFlow \cite{affine}. 
\begin{figure}[]
\centering
\includegraphics[width=.7\textwidth]{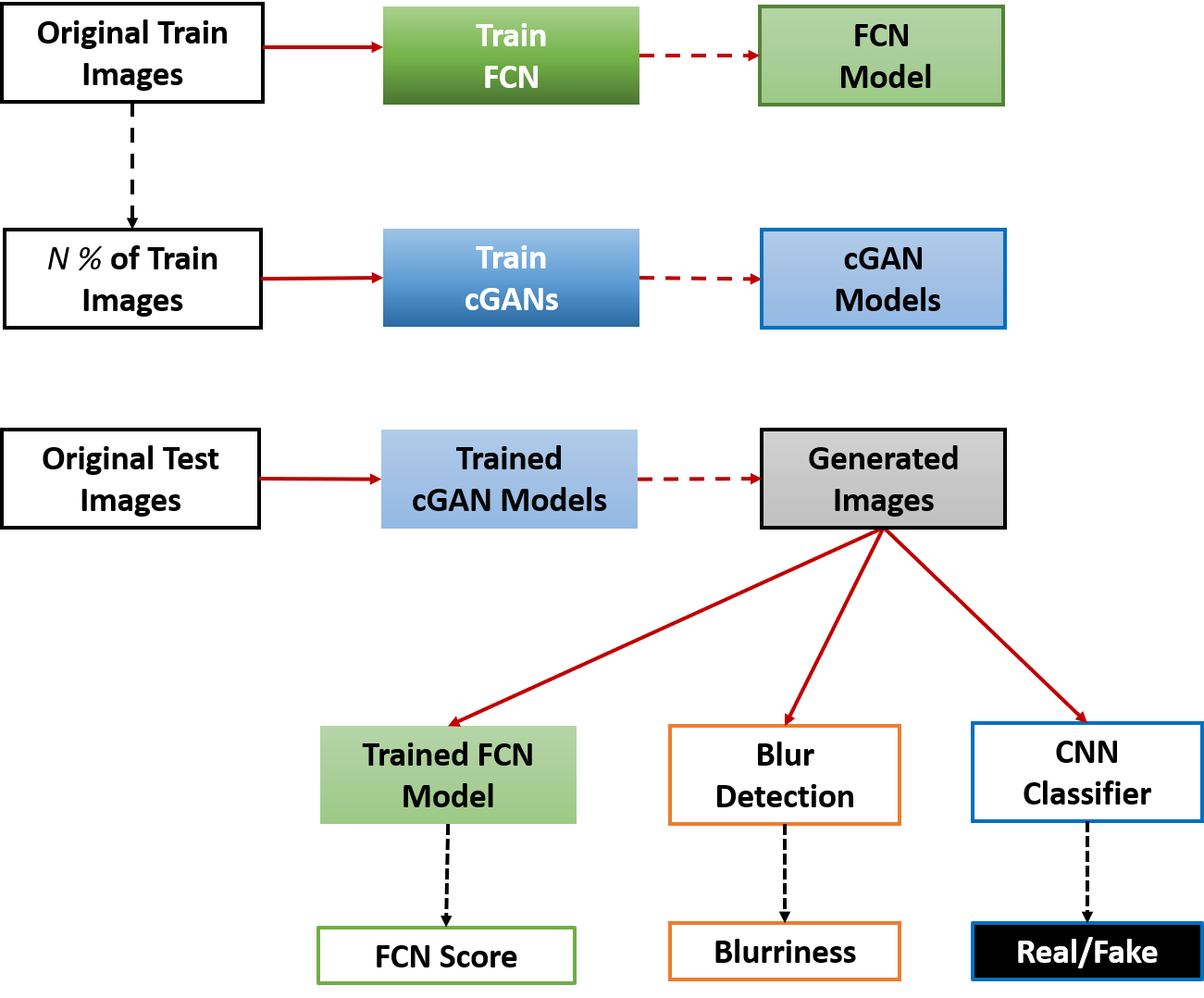}
\caption{Image generation and evaluation scheme. A segmentation model is trained using the full training dataset and generative models are trained with subsets of the original set. Following that, the testing set  is used to evaluate the models.}
\label{metrics}
\end{figure}

The training and evaluation steps are shown in Fig. \ref{metrics}. The \textit{LcGAN} models are trained with \textit{N\%} of the original training data. When selecting the  \textit{N\%} of patients' data, all images from each selected patient  was used.  These images are then evaluated using FCN score, blurriness and the probability value from the CNN classifier. The models that generate the best images based on these metrics are chosen to augment the same \textit{N\%} of data. For the experiments, 2.5\%, 10\% and 25\% of the total number patients  and their CT images were used.  

\subsection{Traditional Augmentation}
\label{sec:tradaug}
\begin{figure}[]
\centering
\includegraphics[width=2.75cm]{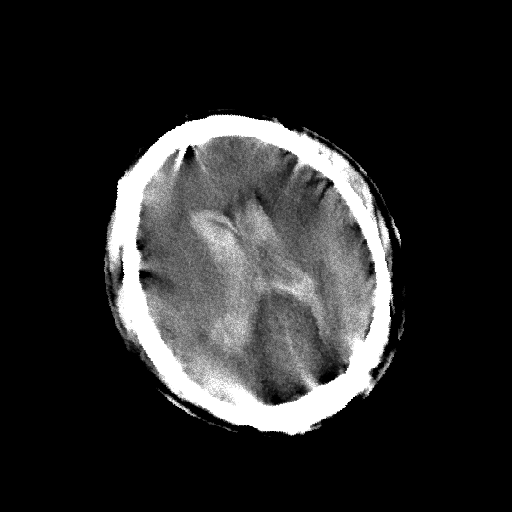}
\includegraphics[width=2.75cm]{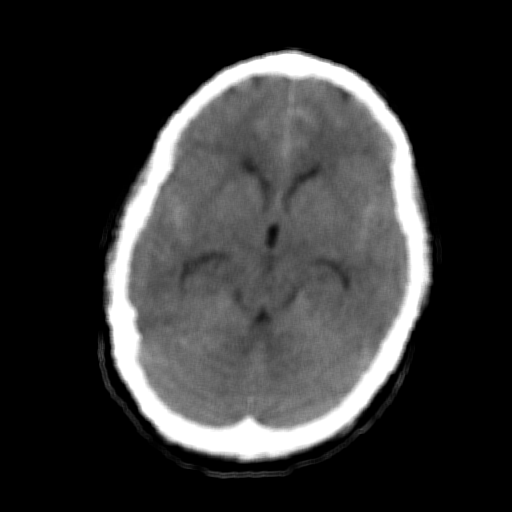}
\includegraphics[width=2.75cm]{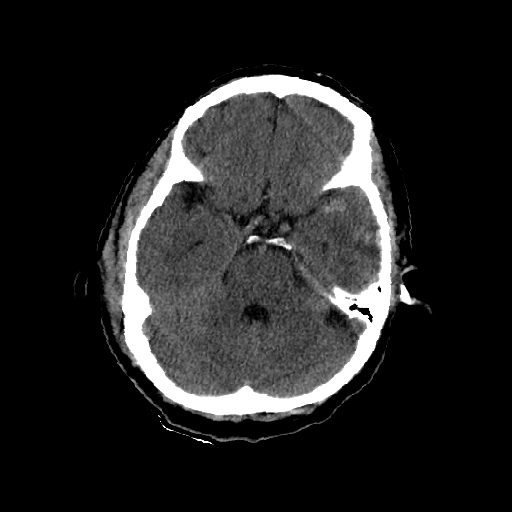}
\includegraphics[width=2.75cm]{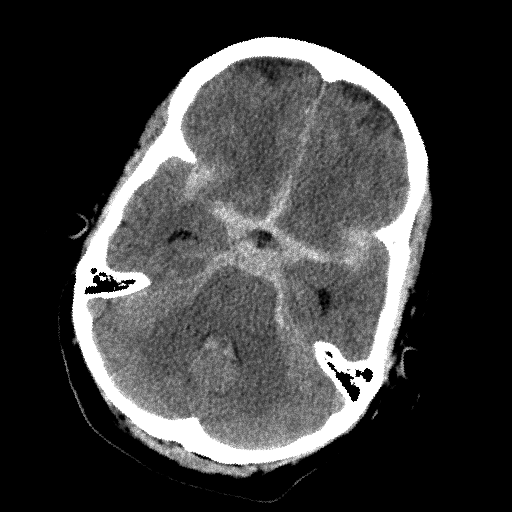}
\caption{Examples with images with different transformations. Each image is transformed with multiple types of augmentation such as rotations, contrast changes, blurring and rescaling. From a specific range of values, random levels are chosen for each transformation.}
\label{trad_aug}
\end{figure}

	Reducing the sizes of data makes the data less diverse and traditional augmentation provides some variance to the dataset. Among the traditional data augmentation methods, only those with techniques that are suitable for medical images were chosen. Affine transformations such as rotation, rescaling, shearing and cropping, as well as pixel level contrast changes and blurring were applied to the selected training images. Fig. \ref{trad_aug} shows some examples of the images after such transformations. In all cases, the addition of these transformations was intended to mimic slight and plausible variations that may occur in such images. Each image was transformed with a certain random level of each transformation and hence another dataset was created of the same size. This dataset was then augmented to the original data during FCN training. Transformation levels for rotation were -25 to 25 degrees,  rescale were -15 to 15 percent increase and  shear were -10 to 10 pixels horizontally. Images were also blurred, and the pixel intensities were changed with a probability of 50\%  for each. In these cases, gaussian blur was applied with radius 2 and for contrast changes, each pixel intensity was increased randomly by up to 50\%. When needed, label images are appropriately transformed so that they match the transformed CT images.

\section{Results and Discussion}
\begin{figure}[]
\centering
\includegraphics[ width = .49\textwidth, height=2.2cm]{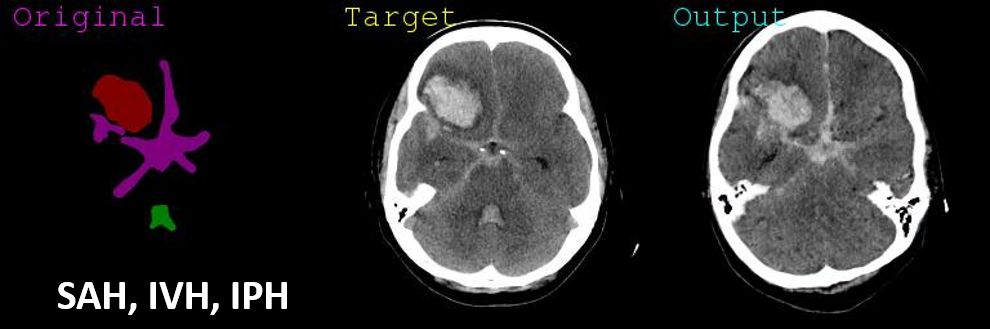}
\includegraphics[ width = .49\textwidth,height=2.2cm]{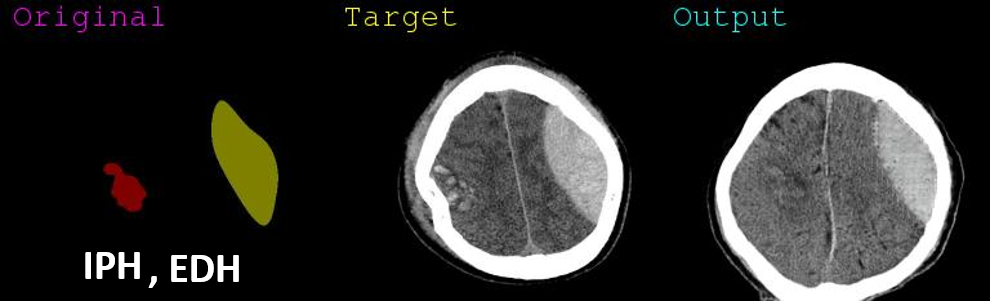}\\
\includegraphics[ width = .49\textwidth,height=2.2cm]{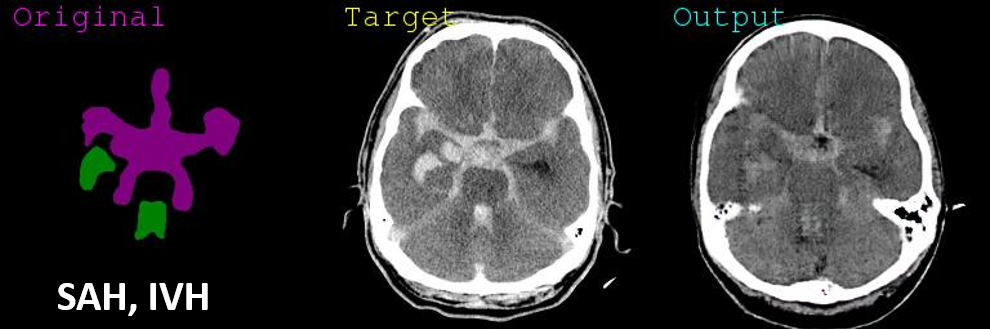}
\includegraphics[ width = .49\textwidth,height=2.2cm]{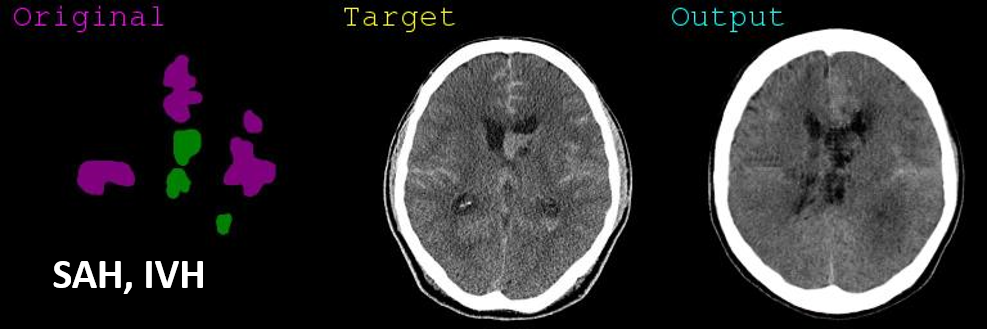}\\
\includegraphics[ width =.49\textwidth,height=2.2cm]{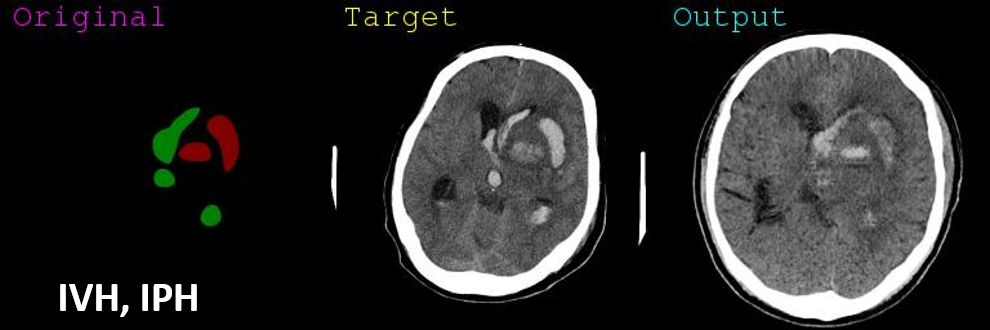}
\includegraphics[ width =.49\textwidth,height=2.2cm]{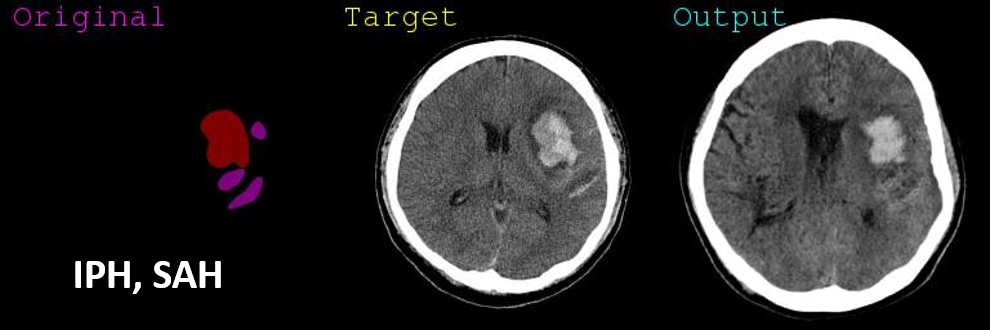}\\
\includegraphics[ width = .49\textwidth,height=2.2cm]{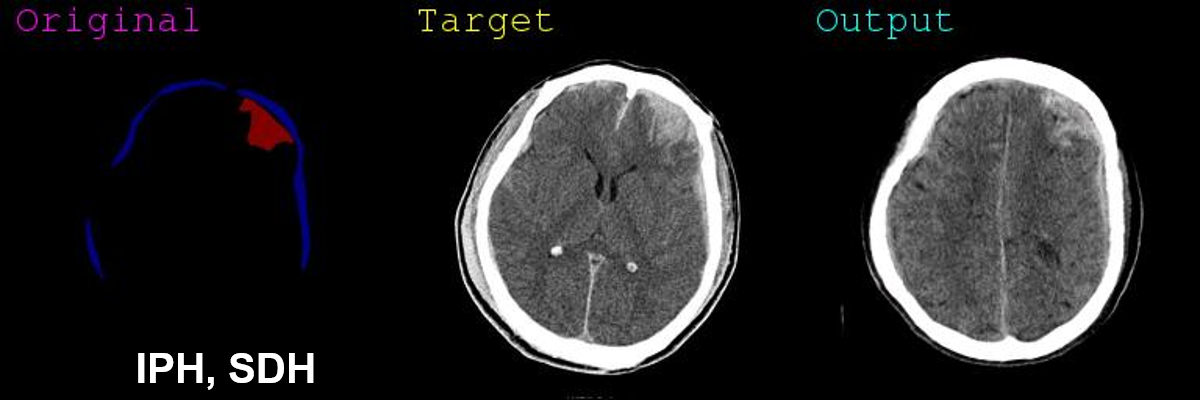}
\includegraphics[ width =.49\textwidth,height=2.2cm]{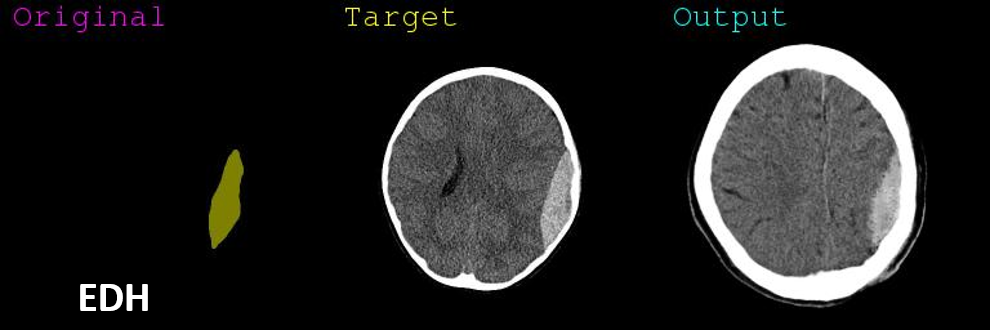}\\
\includegraphics[ width =.49\textwidth,height=2.2cm]{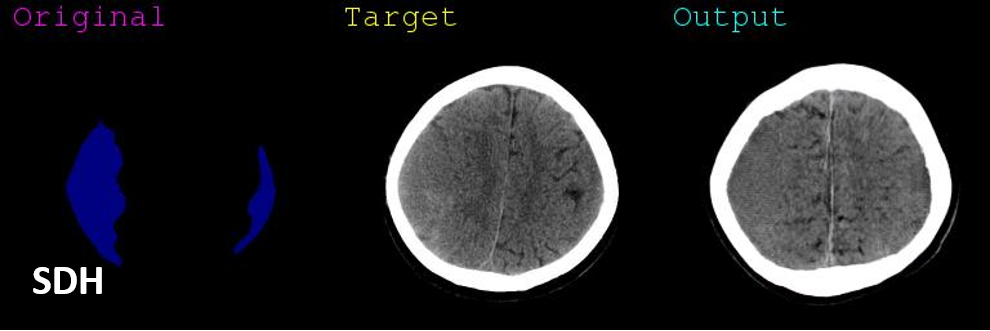}
\includegraphics[ width = .49\textwidth,height=2.2cm]{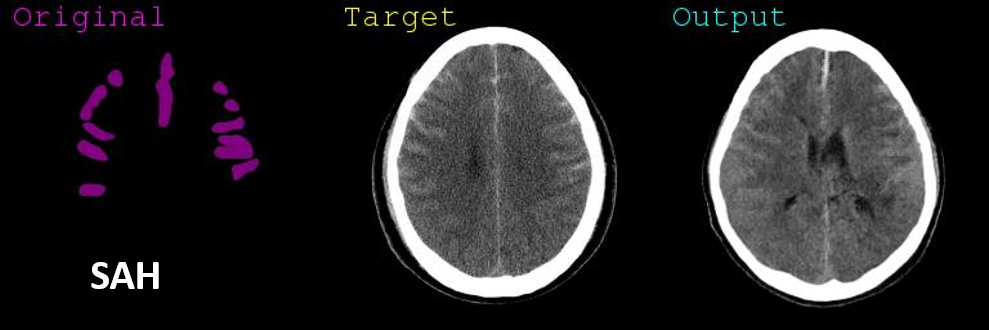}\\
\includegraphics[ width =.49\textwidth,height=2.2cm]{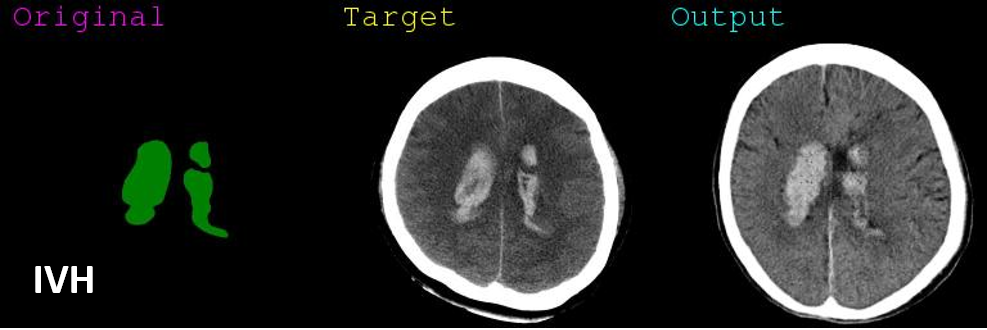}
\includegraphics[ width =.49\textwidth,height=2.2cm]{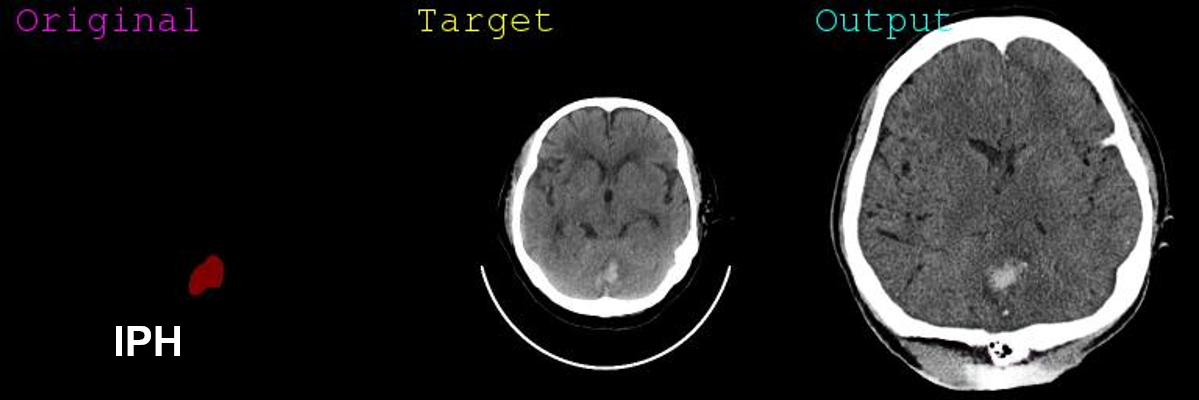}

\caption{Different hemorrhage types and corresponding generated CT images. On each example, a ground truth image (Original) is used as condition to the input CT image (Target) generating  another synthetic CT image (Output).}
\label{gen_aug}
\end{figure}

As mentioned before in the evaluation criteria, background class was ignored during evaluation. Background class consists of large amounts of black pixels all around the brain, and other non-hemorrhage pixels constituting the brain. This allowed a better judgement on performance improvements in the desired classes. Fig. \ref{gen_aug}  shows some examples of generated images (\textit{Output}).  While some of the images are visually like the \textit{Target}, others have some differences.  Some desired variance between the original CT image and synthetic counterpart can be observed. The combination of the data augmentation improved performance in each investigated category. Comparatively high FCN scores and low average blurriness were criteria for good models. Chosen models also had the most number of generated images misclassified as \textit{real} by the CNN classifier. The threshold for CNN classifier was lowered to 10\% (instead of 50\%).
\par
When augmentation was done to the full original dataset, the improvements on each of the metrics was less than 1\%. Adding augmented data (which includes some noisy data) also made the training harder and took significantly more time for large sizes of data. The evaluation metrics defined in Section \ref{sec:eval} were applied to choose the best models trained on each size of data (2.5\%, 10\%, 25\%). As shown in Table. \ref{dsc}, the DSC value increased by 13\% for 2.5\%,  6\% for 10\%  and 2\% when 25\% of data was used. Table \ref{dsc} and \ref{overall} show the performance, on various metrics, of selected number of images and the augmentation by GAN generated images, traditional augmentation and with the addition of both. When 25\% of the original data was used, the DSC segmentation was 3\% worse than using the full (100\%) data. This difference reduced to 2\%  upon augmentation. 

\begin{table}[]
\centering
\caption{DSC Segmentation performance with various amounts of training data}
\label{dsc}
\begin{tabular}{l|ccc}

                               & \textbf{2.5\%}  & \textbf{10\%}  & \textbf{25\%} \\ \hline
\textbf{Selected (Original)}              & .244            & .478           & .565                    \\
\textbf{+ GAN generated}       & .323            & .489           & .571                 \\
\textbf{+ traditional aug.} & .352            & .529           & .580                 \\
\textbf{+ both}                & \textbf{.371}   & \textbf{.538}  & \textbf{.581} \\  \hline
\textbf{Entire dataset (100\%)}    & & \underline{0.628}&  \\
\end{tabular}
\end{table}

\begin{table}[]
\centering
\caption{Overall precision and recall performance with various amounts of training data}
\label{overall}
\begin{tabular}{l|ccc|ccc}
\multicolumn{1}{c|}{}          & \multicolumn{3}{c|}{\textbf{Precision}} & \multicolumn{3}{c}{\textbf{Recall}}   \\ \hline
                               & \textbf{2.5\%}  & \textbf{10\%}  & \textbf{25\%} & \textbf{2.5\%} & \textbf{10\%} & \textbf{25\%} \\ \hline
\textbf{Selected (Original)}              & .408            & .695           & .809          & .360           & .681          & .747          \\
\textbf{+ GAN generated}       & .519            & .719           & .810          & .479           & .684          & .757          \\
\textbf{+ traditional aug.} & .549            & .762           & .834          & .502           & .724          & .763          \\
\textbf{+ both}                & \textbf{.607}   & \textbf{.766}  & \textbf{.844} & \textbf{.540}  & \textbf{.737} & \textbf{.777} \\  \hline
\textbf{Entire dataset (100\%)}    & & \underline{0.850} & & & \underline{0.802} &  \\
\end{tabular}
\end{table}
\begin{table}[]
\centering
\caption{Segmentation performance (DSC) for each class using various \% of training data}
\begin{tabular}{l|ccccc|c}
                                & \textbf{IPH} & \textbf{IVH} & \textbf{EDH} & \textbf{SDH} & \textbf{SAH} & \textbf{Mean} \\ \hline
\textbf{2.5\%} (Original)                & .772         & .186         & .424         & .558        &.246        & .427          \\
\textbf{+ GAN generated}        &.773         & .513         & .672         & .602        & .456         & .603         \\
\textbf{+ traditional augment}  &\textbf{.782}        & \textbf{.553}         &.612        &.635        & .481         & .613          \\
\textbf{+ both}                 &.777        &.517       &\textbf{.708}         & \textbf{.642}         & \textbf{.529}        &\textbf{.635}          \\ \hline
\textbf{10\%} (Original)                & .809        & .682         & .716         & .690        &.629        & .705          \\
\textbf{+ GAN generated}        &.816         & .685         & .738        & .691        & .630         & .712         \\
\textbf{+ traditional augment}  &.818        & .702         &.768        &.737        &\textbf{.645}         & .734          \\
\textbf{+ both}                 &\textbf{.821}     &\textbf{.715}       &\textbf{.775}         & \textbf{.752}         &.636        &\textbf{.740}          \\ \hline
\textbf{25\%} (Original)                  & .829         & .737         & .811         & .767         &\textbf{.654}        & .760          \\
\textbf{+ GAN generated}        &\textbf{.832}         & .748         & .832         & .766         & .652         & .766          \\
\textbf{+ traditional augment}  & .828         & .737         &\textbf{.837}        &\textbf{.781}        & .652         & .767          \\
\textbf{+ both}                 &\textbf{.832}         &\textbf{.750}         &\textbf{.837}         & .780         & .653        &\textbf{.771}          \\ \hline
\textbf{100\% (Entire Dataset)} &\underline{.842}         & \underline{.760}         &\underline{.872}         & \underline{.817}         &\underline{.657}        &\underline{.789}         
\end{tabular}
\label{per_class}
\end{table}

\begin{figure}[]
\centering
\includegraphics[width=.9\textwidth]{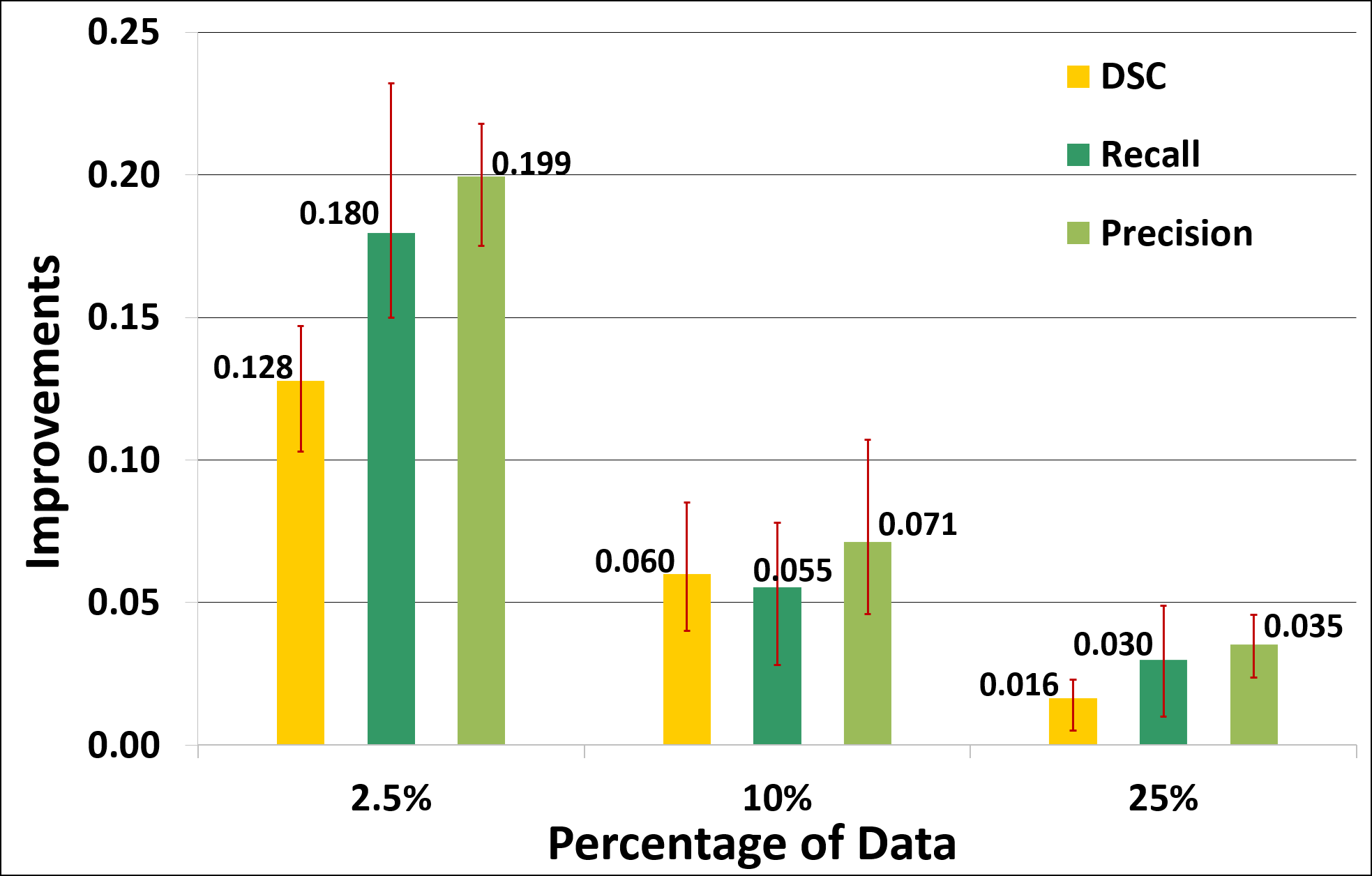}
\caption{DSC, recall and precision improvement with various percentages of data with combined augmentation}
\label{improve}
\end{figure}

Performance of each of the classes improved mostly with augmentation. However, in some cases the performance of segmentation on some classes was slightly worse but significantly better on others as shown in Table \ref{per_class}. The DSC score for segmentation is calculated without any thresholds. This is the single performance criteria that was used to monitor the progress of training the segmentation model. Fig. \ref{improve} shows that as the number of data increased the effect of augmentation on performance decreased. When separately compared, traditional augmentation had better performance than the generative approach. But in almost all comparisons, the combination of both had the best performance.  Adding traditional augmentation to the generative algorithm training did not yield any overall improvements.

\section{Conclusion and Future Work}
We present a conditionally generative adversarial algorithm to enhance the segmentation performance of hemorrhagic lesions in CT images. Training with a condition based on the ground truth empirical image creates a data-label pair of images, which is ready for augmentation. In conclusion, performance of segmentation models trained with smaller sizes of data benefited more from the addition of synthetic data. The difference in performance between the constrained and the full datasets was reduced considerably when the traditional and GAN generated augmentation methods were combined.  Further refinement can be done in training the generative network to be able to quantify the quality of images. Ground truth labels that are considered to be plausible by experts; but not labeled from existing CT hemorrhages could be used to generate more CT images. Also, an end-to-end system could simplify the training and reduce disk space.
\bibliography{JDI_v4}

\end{document}